\pdfoutput=1
\RequirePackage{ifpdf}
\ifpdf % We are running pdfTeX in pdf mode
\documentclass[pdftex]{sigma}
\else
\documentclass{sigma}
\fi

\usepackage{mathrsfs}

\numberwithin{equation}{section}

\newtheorem{Theorem}{Theorem}[section]

\begin{document}

%\allowdisplaybreaks

\newcommand{\arXivNumber}{1605.07770}

\renewcommand{\PaperNumber}{091}

\FirstPageHeading

\ShortArticleName{Recursion Operators and Tri-Hamiltonian Structure}

\ArticleName{Recursion Operators and Tri-Hamiltonian Structure\\ of the First Heavenly Equation of Pleba\'nski}

\Author{Mikhail B. SHEFTEL~$^\dag$ and Devrim YAZICI~$^\ddag$}

\AuthorNameForHeading{M.B.~Sheftel and D.~Yaz{\i}c{\i}}

\Address{$^\dag$~Department of Physics, Bo\u{g}azi\c{c}i University, Bebek, 34342 Istanbul, Turkey}
\EmailD{\href{mailto:mikhail.sheftel@boun.edu.tr}{mikhail.sheftel@boun.edu.tr}}
\URLaddressD{\url{http://www.phys.boun.edu.tr/faculty_wp/mikhail_sheftel.html}}

\Address{$^\ddag$~Department of Physics, Y{\i}ld{\i}z Technical University, Esenler, 34220 Istanbul, Turkey}
\EmailD{\href{mailto:yazici@yildiz.edu.tr}{yazici@yildiz.edu.tr}}

\ArticleDates{Received June 28, 2016, in f\/inal form September 10, 2016; Published online September 14, 2016}

\Abstract{We present f\/irst heavenly equation of Pleba\'nski in a two-component evolu\-tio\-nary form and obtain Lagrangian and Hamiltonian representations of this system. We study all point symmetries of the two-component system and, using the inverse Noether theorem in the Hamiltonian form, obtain all the integrals of motion corresponding to each variational (Noether) symmetry. We derive two linearly independent recursion operators for symmetries of this system related by a discrete symmetry of both the two-component system and its symmetry condition. Acting by these operators on the f\/irst Hamiltonian operator~$J_0$ we obtain second and third Hamiltonian operators. However, we were not able to f\/ind Hamiltonian densities corresponding to the latter two operators. Therefore, we construct two recursion operators, which are either even or odd, respectively, under the above-mentioned discrete symmetry. Acting with them on~$J_0$, we generate another two Hamiltonian operators~$J_+$ and~$J_-$ and f\/ind the corresponding Hamiltonian densities, thus obtaining second and third Hamiltonian representations for the f\/irst heavenly equation in a~two-component form. Using P.~Olver's theory of the functional multi-vectors, we check that the linear combination of~$J_0$,~$J_+$ and~$J_-$ with arbitrary constant coef\/f\/icients satisf\/ies Jacobi identities. Since their skew symmetry is obvious, these three operators are compatible Hamiltonian operators and hence we obtain a tri-Hamiltonian representation of the f\/irst heavenly equation. Our well-founded conjecture applied here is that P.~Olver's method works f\/ine for nonlocal operators and our proof of the Jacobi identities and bi-Hamiltonian structures crucially depends on the validity of this conjecture.}


\Classification{35Q75; 83C15; 37K05; 37K10}

\section{Introduction}

Pleba\'nski had demonstrated how the Einstein f\/ield equations in the complex domain under (anti-)self-duality condition reduce to a single scalar PDE,
which he called the f\/irst and the second heavenly equations \cite{pleb}. Earlier we had shown that the second heavenly equation is a~bi-Hamiltonian system~\cite{nns}. We had also obtained bi-Hamiltonian representations for all other equations of the heavenly type~\cite{nsky,smy,sy,Ya} except the f\/irst heavenly equation.

Here we show that the f\/irst heavenly equation of Pleba\'nski
\begin{gather}
u_{\tilde{t}\tilde{y}}u_{\tilde{x}\tilde{z}} - u_{\tilde{t}\tilde{z}}u_{\tilde{x}\tilde{y}} = 1.
\label{heav1}
\end{gather}
presented in the two-component evolutionary form
\begin{gather}
 u_t = v,\nonumber\\
v_t = u_{xx} + \frac{1}{\displaystyle u_{\tilde{x}\tilde{z}}}\{(v_{\tilde{z}} + u_{x\tilde{z}})(v_{\tilde{x}} - u_{x\tilde{x}}) + 1\} \equiv Q,
\label{2comp}
\end{gather}
where $t = \tilde{t} + \tilde{y}$, $x=\tilde{t} - \tilde{y}$, possesses two linearly independent recursion operators for symmetries. (From now on, letter subscripts denote partial derivatives with respect to corresponding variables.) It should be noted here that recursion operators and Hamiltonian structures can also be found in the initial, non-evolutionary form of a given PDE by the methods of~\cite{KVV}.

In Section~\ref{Hamilt}, we present the Lagrangian for the system \eqref{2comp}. Lagrangian of this system turns out to be degenerate and therefore we use the Dirac's theory of constraints~\cite{dirac}. The Poisson bracket of Dirac constraints provides us a symplectic operator and symplectic 2-form. The inverse of the symplectic operator is the f\/irst Hamiltonian operator $J_0$ for the two-component system. We f\/ind the corresponding Hamiltonian density and present our system in a Hamiltonian form.

In Section~\ref{sym_int}, we obtain all point Lie symmetries of the f\/irst heavenly system (FHS) together with the table of commutators of the basis Lie algebra generators and show how the correspon\-ding Hamiltonians of the variational symmetry f\/lows can be obtained from the inverse Noether theorem in a Hamiltonian form. These Hamiltonians are integrals of the motion along the f\/irst heavenly f\/low for all the variational symmetry f\/lows that commute with FHS. Each Hamiltonian is also conserved along all the symmetry f\/lows that commute with the f\/low generated by this Hamiltonian.

In Section \ref{recurs}, we show that the symmetry condition for the f\/irst heavenly equation~\eqref{heav1} in a one component form can be compactly expressed in terms of Lax operators and have a two dimensional total divergence form and therefore def\/ines a potential variable. We prove that the potential of a~symmetry is again a symmetry and thereby a recursion relation for symmetries is derived. These are partner symmetries which we introduced in~\cite{mns} for the complex Monge--Amp\`ere equation and later classif\/ied heavenly equations possessing partner symmetries in~\cite{shma}, with particular applications in~\cite{nns} for the Pleba\'nski second heavenly equation and in~\cite{sy} for the mixed heavenly and Husain systems. In~\cite{mash,Sh-Ma} we have demonstrated how to use partner symmetries for constructing noninvariant solutions of the complex Monge--Amp\`ere equation and corresponding (anti-)self-dual Ricci-f\/lat metrics without Killing vectors which are important for discovering an explicit metric of the gravitational instanton $K3$.

We have obtained the recursion operator $R_1$ for symmetries in a two-component form. We also discover a second recursion operator $R_2$ by applying to $R_1$
a discrete symmetry of the system~\eqref{2comp} and its symmetry condition. We also construct the recursion operators $R_+ = R_1 + R_2$ and $R_- = R_1 - R_2$
that are even and odd, respectively, under the discrete symmetry. The idea of constructing recursion operators from the Lax-type representation in the context of partner symmetries appeared in our paper~\cite{mns} and later, in a more general context, in the paper~\cite{Artur}. %, version 1 (2015).
 Here we derive recursion operators using our approach of partner symmetries while A.~Sergyeyev in~\cite{Artur} uses a more general approach for deriving recursion operators which works even when partner symmetries do not exist (see, e.g., our paper~\cite{smy}).

In Section~\ref{bi-Hamilt}, acting by these recursion operators on $J_0$ we obtain several alternatives for the second Hamiltonian operator for the two-component system, $J_1=R_1J_0$, $J_2=R_2J_0$ and $J_\varepsilon = R_\varepsilon J_0$, where $\varepsilon = \pm$. We have failed to f\/ind the Hamiltonian densities for $J_1$ and $J_2$ such that provide bi-Hamiltonian representations of the two-component system. On the contrary, for the even and odd Hamiltonian operators $J_\varepsilon$ we have found the appropriate Hamiltonian densities and obtained two bi-Hamiltonian representations of the f\/irst heavenly equation of Pleba\'nski in a two-component form~\eqref{2comp}. Using jointly all three Hamiltonian operators $J_0$, $J_+$ and $J_-$ we obtain \textit{tri-Hamiltonian} representation of the system~\eqref{2comp}, provided that the Poisson compatibility of these operators is proved.

In Section \ref{Jacobi}, we prove that the Jacobi identities are satisf\/ied for an arbitrary linear combination of the three operators and hereby prove that they are indeed Hamiltonian and compatible. For this purpose we use P.~Olver's theory of functional multi-vectors~\cite{olv}. P.~Olver's theory was formulated for matrix-dif\/ferential Hamiltonian operators while our operators $J_\varepsilon$ are nonlocal. We show how this theory works nicely even for these nonlocal operators, so that the applicability of P.~Olver's method to nonlocal Hamiltonian operators seems to be a well-founded conjecture though a rigorous generalization of his theory to nonlocal Hamiltonian operators is still awaited. Under this conjecture, we have shown that equations~\eqref{2comp} yield a tri-Hamiltonian completely integrable system in the sense of Magri~\cite{magri,magri+}.

It should be noted that there is a number of tri-Hamiltonian systems (and even multi-Hamiltonian systems) in the literature \cite{AF87,AF88,AF89} and~\cite{Kuper}. Our results show that a tri-Hamiltonian system can be found also in the multi-dimensional case.

We also note that the Olver's method for for checking Jacobi identity is not the only one that works for nonlocal operators. Many authors use the formula by Gel'fand, Dorfman and co-workers. Recent examples of $(1{+}1)$-dimensional systems treated in this way can be found in~\cite{LZ}.

\section{Lagrangian, symplectic and Hamiltonian structures}\label{Hamilt}

We obtain a Lagrangian for this two-component system
\begin{gather}
L = \left(vu_t - \frac{v^2}{2}\right)u_{\tilde{x}\tilde{z}} + u_t\left(\frac{1}{3}u_xu_{\tilde{x}\tilde{z}} + \frac{2}{3}u_{\tilde{z}}u_{\tilde{x}x}\right)
- \frac{1}{2}u_x^2 u_{\tilde{x}\tilde{z}} + u,
\label{L}
\end{gather}
which is degenerate in the following sense. The canonical momenta
\begin{gather}
\Pi_1=\Pi_u = \frac{\partial L}{\partial u_t} = vu_{\tilde{x}\tilde{z}} + \frac{1}{3} (u_{\tilde{x}}u_{\tilde{x}\tilde{z}} + 2u_{\tilde{z}}u_{\tilde{x}x}),\qquad
\Pi_2=\Pi_v = \frac{\partial L}{\partial v_t} = 0,
\label{mom}
\end{gather}
which satisfy canonical Poisson brackets, cannot be inverted for velocities. Following Dirac's theory of constraints \cite{dirac}, we impose \eqref{mom}
as constraints
\begin{gather}
 \Phi_1=\Phi_u = \Pi_u - vu_{\tilde{x}\tilde{z}} - \frac{1}{3} (u_{\tilde{x}}u_{\tilde{x}\tilde{z}} + 2u_{\tilde{z}}u_{\tilde{x}x}) = 0,\qquad
 \Phi_2=\Phi_v = \Pi_v = 0,\label{constr}
\end{gather}
and calculate the Poisson bracket of the left-hand sides of the constraints \eqref{constr}
\begin{gather*} K_{ij}=\big[\Phi_i(x,\tilde{x},\tilde{z}),\Phi_j(x',\tilde{x}',\tilde{z}')\big]. \end{gather*}
The result is expressed in the matrix-operator form as
\begin{gather}
K = \left(
\begin{matrix}
(v_{\tilde{z}}+u_{x\tilde{z}})D_{\tilde{x}} + (v_{\tilde{x}}-u_{x\tilde{x}})D_{\tilde{z}} + v_{\tilde{x}\tilde{z}} & - u_{\tilde{x}\tilde{z}}\\
u_{\tilde{x}\tilde{z}} & 0
\end{matrix}\right),\label{K}
\end{gather}
which is obviously skew symmetric. From now on $D_\xi$ denotes total derivative operator with respect to $\xi$. The corresponding 2-form $\Omega$ is obtained by integrating the density
\begin{gather*}
\omega = \frac{1}{2}\left\{\right(v_{\tilde{z}}+u_{x\tilde{z}})du\wedge du_{\tilde{x}} + (v_{\tilde{x}}-u_{x\tilde{x}})du\wedge du_{\tilde{z}}\}
+ u_{\tilde{x}\tilde{z}}dv\wedge du ,
%\label{twoform}
\end{gather*}
which is closed up to a total divergence
\begin{gather*}
d\omega = du_x\wedge du_{\tilde{x}}\wedge du_{\tilde{z}}
= \frac{1}{3}\big\{
D_x(du\wedge du_{\tilde{x}}\wedge du_{\tilde{z}}) + D_{\tilde{x}}(du\wedge du_x\wedge du_{\tilde{z}})\\
 \hphantom{d\omega = du_x\wedge du_{\tilde{x}}\wedge du_{\tilde{z}}=}{} + D_{\tilde{z}}(du_x\wedge du_{\tilde{x}}\wedge du) \big\},
\end{gather*}
so that $\Omega = \iiint_V \omega dV$ is closed under appropriate boundary conditions and hereby $\Omega$ is a~symplectic form, while $K$ in \eqref{K} is a symplectic operator. Taking the inverse of $K$ we obtain Hamiltonian operator
\begin{gather}
J_0 = K^{-1} = \left(\begin{matrix}
 0 &\displaystyle \frac{1}{u_{\tilde{x}\tilde{z}}} \\
 - \displaystyle \frac{1}{u_{\tilde{x}\tilde{z}}} &\displaystyle \frac{1}{u_{\tilde{x}\tilde{z}}} \{(v_{\tilde{z}}+u_{x\tilde{z}})D_{\tilde{x}} + (v_{\tilde{x}}-u_{x\tilde{x}})D_{\tilde{z}} + v_{\tilde{x}\tilde{z}}\} \frac{1}{u_{\tilde{x}\tilde{z}}}
 \end{matrix}\right),\label{J0}
\end{gather}
which is obviously skew symmetric, while the Jacobi identities for $J_0$ are satisf\/ied as a~consequence of the closure of the symplectic 2-form $\Omega$.
The corresponding Hamiltonian density $H_1$ is found by using \eqref{L} and \eqref{mom} in the relation $H_1 = \Pi_u u_t + \Pi_v v_t - L$ which yields
\begin{gather}
 H_1 = \frac{1}{2}\big(v^2 + u_x^2\big)u_{\tilde{x}\tilde{z}} - u.
 \label{H1}
\end{gather}
Thus, we present the two-component system \eqref{2comp} in the Hamiltonian form
\begin{gather}\label{J0H1}
 \left( \begin{matrix}
 u_t\\ v_t
 \end{matrix}\right) = J_0
 \left( \begin{matrix}
 \delta_uH_1\\ \delta_vH_1
 \end{matrix}\right)
\end{gather}
where $\delta_u$, $\delta_v$ denote Euler--Lagrange operators related to variational derivatives of Hamiltonian functional~\cite{olv}.

\section{Symmetries and integrals of motion}\label{sym_int}

Point Lie symmetries of the system \eqref{2comp} are determined by the following basis generators
\begin{gather}
 X_1 = \partial_{\tilde{z}},\quad X_2 = \partial_{\tilde{x}},\qquad X_3 = \tilde{z}\partial_{\tilde{z}} - \tilde{x}\partial_{\tilde{x}},
 \qquad X_4 = t\partial_t + x\partial_x + u\partial_u, \nonumber\\
 X_5 = 2\tilde{z}\partial_{\tilde{z}} + u\partial_u + v\partial_v,\qquad Y_a = a(\tilde{x})(\partial_t + \partial_x),
 \quad Z_b = b(\tilde{z})(\partial_x - \partial_t),\nonumber \\
 V_{f,g} = \{f(t+x,\tilde{x}) + g(t-x,\tilde{z})\}\partial_u + \{f_t(t+x,\tilde{x}) + g_t(t-x,\tilde{z})\}\partial_v,\label{symm}
\end{gather}
where $a$, $b$ and $f$, $g$ are arbitrary smooth functions of one and two variables, respectively and subscripts denote partial derivatives. We note that the obvious translation invariance symmetries are generated by $X_1$, $X_2$ and also by $Y_{a=1}\pm Z_{b=1}$. In particular, $X = (Z_{b=1}-Y_{a=1})/2 = - \partial_t$ is the generator of the f\/irst heavenly f\/low~\eqref{2comp} itself.

\begin{table}[ht]\centering\caption{Commutators of point symmetries of the f\/irst heavenly system.}\label{Tab1}\vspace{1mm}
\begin{tabular}{|c|c|c|c|c|c|c|c|c|}
\hline &$X_1$&$X_2$ &$X_3$& $X_4$ &$X_5$ &$Y_a$ &$Z_b$ &$V_{f,g}$\bsep{2pt}
\\ \hline
 $X_1$ & $0$ & $0$& $X_1$ & $0$ &$2X_1$ &$0$ &$Z_{b'}$ &$V_{0,g_{\tilde{z}}}$\bsep{2pt}
\\ \hline
 $X_2$ &$0$& $0$ & $-X_2$ & $0$ &$0$&$Y_{a'}$& $0$& $V_{f_{\tilde{x}},0}$\bsep{2pt}
\\ \hline
 $X_3$ &$-X_1$ &$X_2$ & $ 0$ &$0$ &$0$& $-Y_{\tilde{x}a'}$& $Z_{\tilde{z}b'}$& $V_{-\tilde{f},\tilde{g}}$\bsep{2pt}
\\ \hline
 $X_4$ &$0$& $0$& $0$ & $0$& $0$& $0$& $0$ & $V_{\hat{f},\hat{g}}$\bsep{2pt}
\\ \hline
 $X_5$ &$-2X_1$& $0$& $0$& $0$& $0$& $0$& $2Z_{\tilde{z}b'}$& $V_{-f,2\tilde{g}-g}$\bsep{2pt}
\\ \hline
$Y_a$ &$0$& $-Y_{a'}$& $Y_{\tilde{x}a'}$& $0$& $0$& $0$& $0$&$V_{2af_t,0}$\bsep{2pt}
\\ \hline
$Z_b$ &$-Z_{b'}$& $0$& $-Z_{\tilde{z}b'}$& $0$& $-2Z_{\tilde{z}b'}$& $0$& $0$& $V_{0,-2bg_t}$\bsep{2pt}
\\ \hline
 $V_{f,g}$ &$V_{0,-g_{\tilde{z}}}$ &$V_{-f_{\tilde{x}},0}$ & $V_{\tilde{f},-\tilde{g}}$& $V_{-\hat{f},-\hat{g}}$& $V_{f,g-2\tilde{g}}$ &$V_{-2af_t,0}$
 &$V_{0,2bg_t}$& $0$\bsep{2pt}
\\ \hline
\end{tabular}
\end{table}

The Lie algebra of point symmetries is determined by the Table~\ref{Tab1} of commutators of the basis generators where the commutator $[X_i,X_j]$ stands at the intersection of $i$th row and $j$th column. Here we use the following shorthand notation{\samepage
\begin{gather*}\tilde{f}=\tilde{x}f_{\tilde{x}},\qquad \tilde{g}=\tilde{z}g_{\tilde{z}},\qquad \hat{f}=(t+x)f_t-f,\qquad\hat{g}=(t-x)g_t-g,
\end{gather*}
and the primes denote derivatives of arbitrary functions $a$ and $b$ of a single variable.}

We need symmetry characteristics determining symmetries in evolutionary form \cite{olv} with the independent variables not being transformed under symmetry transformations. For the point symmetry generator of the form $X = \xi^i\partial_{x^i} + \eta^\alpha\partial_{u^\alpha}$, where the summation over repeated indices is used, the symmetry characteristics are def\/ined as $\varphi^\alpha = \eta^\alpha - u^\alpha_i\xi^i$ with the subscripts~$i$ denoting derivatives with respect to $x^i$. In our problem, $\alpha = 1,2$, $u^1=u, u^2=v$, $\eta^1=\eta^u$, $\eta^2=\eta^v$, $\sum_i = \sum\limits_{i=1}^4$, $x^1 = t$, $x^2 = x$, $x^3 = \tilde{x}$, $x^4 = \tilde{z}$ and $\varphi^1 = \varphi$ while $\varphi^2 = \psi$, where $\varphi$ and $\psi$ determine the transformation of $u$ and $v$, respectively. We also use $u_t = v$ and $v_t =Q$ where $Q$ is the right-hand side of the second of our equations~\eqref{2comp}. Symmetry characteristics have the form
\begin{gather}
 \varphi = \eta^u - v\xi^t - u_x\xi^x - u_{\tilde{x}}\xi^{\tilde{x}} - u_{\tilde{z}}\xi^{\tilde{z}}, \qquad
 \psi = \eta^v - Q\xi^t - v_x\xi^x - v_{\tilde{x}}\xi^{\tilde{x}} - v_{\tilde{z}}\xi^{\tilde{z}} .
\label{char}
\end{gather}
Applying the formula \eqref{char} to the generators in \eqref{symm}, we obtain characteristics of these sym\-met\-ries
\begin{gather}
 \varphi_1 = - u_{\tilde{z}},\qquad \psi_1 = - v_{\tilde{z}}, \qquad \varphi_2 = - u_{\tilde{x}},\qquad \psi_2 = - v_{\tilde{x}}, \nonumber \\
 \varphi_3 = \tilde{x}u_{\tilde{x}} - \tilde{z}u_{\tilde{z}},\qquad \psi_3 = \tilde{x}v_{\tilde{x}} - \tilde{z}v_{\tilde{z}}, \qquad
 \varphi_4 = u - tv - xu_x,\qquad \psi_4 = - tQ - xv_x, \nonumber \\
 \varphi_5 = u - 2\tilde{z}u_{\tilde{z}},\qquad \psi_5 = v - 2\tilde{z}v_{\tilde{z}},\qquad \varphi_a = -a(\tilde{x})(v+u_x),
 \nonumber\\
 \psi_a = -a(\tilde{x})(Q+v_x),\qquad \varphi_b = b(\tilde{z})(v-u_x),\qquad \psi_b = b(\tilde{z})(Q-v_x), \nonumber\\
 \varphi_{f,g} = f(t+x,\tilde{x}) + g(t-x,\tilde{z}),\qquad \psi_{f,g} = f_t(t+x,\tilde{x}) + g_t(t-x,\tilde{z}). \label{chartab}
\end{gather}
In accordance with our previous remark that $X = (Z_{b=1}-Y_{a=1})/2 =-\partial_t$ generates the f\/irst heavenly f\/low, we note that $(\varphi_{b=1} - \varphi_{a=1})/2 = v$ and $(\psi_{b=1} - \psi_{a=1})/2 = Q$ are characteristics of this symmetry with the group parameter $\tau = t$.

First Hamiltonian structure provides a link between symmetries in evolutionary form and integrals of motion conserved by the Hamiltonian f\/low~\eqref{J0H1}. Replacing time $t$ by the group parameter~$\tau$ in~\eqref{J0H1} and using $u_\tau = \varphi$, $v_\tau = \psi$ for symmetries in the evolutionary form, we obtain the Hamiltonian form of the Noether theorem for any conserved density $H$ of an integral of motion
\begin{gather}
 \left(\begin{matrix} \varphi
 \\ \psi
 \end{matrix}
 \right) = J_0 \left(\begin{matrix}
 \delta_u H \\ \delta_v H \end{matrix}
 \right).
\label{Noether}
\end{gather}
To determine the integral $H$ that corresponds to a known symmetry with the characteristic $(\varphi, \psi)$ we use the inverse Noether theorem
\begin{gather}
 \left(\begin{matrix}
 \delta_u H \\ \delta_v H \end{matrix}
 \right) = K
 \left(\begin{matrix} \varphi
 \\ \psi
 \end{matrix}
 \right),\label{InvNoeth}
\end{gather}
where operator $K = J_0^{-1}$ is def\/ined in \eqref{K}. Here \eqref{InvNoeth} is obtained by applying the operator $K$ to both sides of~\eqref{Noether}.

We now apply the formula \eqref{InvNoeth} to determine integrals $H^i$ corresponding to all variational symmetries with characteristics $(\varphi_i, \psi_i)$
from the list~\eqref{chartab}. Using the expression~\eqref{K} for~$K$, we rewrite the formula~\eqref{InvNoeth} in an explicit form
\begin{gather*}
 \left(\begin{matrix}
 \delta_u H^i \\ \delta_v H^i \end{matrix}
 \right) = \left(\begin{matrix}
 K_{11} & - u_{\tilde{x}\tilde{z}} \\
 u_{\tilde{x}\tilde{z}} & 0
 \end{matrix}
 \right) \left(\begin{matrix} \varphi_i
 \\ \psi_i
 \end{matrix}
 \right),%\label{explicit}
\end{gather*}
which supplies the formulas for determining integrals $H^i$ for the known symmetries $(\varphi_i, \psi_i)$
\begin{gather}
 \delta_u H^i = K_{11}\varphi_i - u_{\tilde{x}\tilde{z}}\psi_i,\qquad \delta_v H^i = u_{\tilde{x}\tilde{z}}\varphi_i .
\label{sym_H}
\end{gather}
Here $K_{11}$ is determined from \eqref{K} to be $K_{11} = (v_{\tilde{z}}+u_{x\tilde{z}})D_{\tilde{x}} + (v_{\tilde{x}}-u_{x\tilde{x}})D_{\tilde{z}} + v_{\tilde{x}\tilde{z}}$. We always start with solving the second equation in \eqref{sym_H} in which, due to the fact that $\varphi_i$ never contain derivatives of $v$, $\delta_v H^i$ is reduced to the partial derivative $H^i_v$, so that the equation $H^i_v = u_{\tilde{x}\tilde{z}}\varphi_i$ is easily integrated with respect to $v$ with the ``constant of integration'' $h^i[u]$ depending only on~$u$ and its derivatives. Then the operator $\delta_u$ is applied to the resulting $H^i$, which involves the unknown $\delta_u h^i[u]$, and then it is equated to $\delta_u H^i$ following from the f\/irst equation in~\eqref{sym_H} to determine $\delta_u h^i[u]$. Finally, we reconstruct $h^i[u]$ and hence $H^i$. If we encounter a contradiction, then this particular symmetry is not a variational one and does not lead to an integral.

This solution algorithm for the symmetries listed in \eqref{symm} with characteristics \eqref{chartab} yields the following results. $X_5$ generates a non-variational symmetry. For all other symmetries the corresponding integrals read
\begin{gather}
 H^1 = - \left(vu_{\tilde{z}}u_{\tilde{x}\tilde{z}} + \frac{1}{2} u_{\tilde{z}}^2 u_{x\tilde{x}}\right),\qquad
 H^2 = - \left(vu_{\tilde{x}}u_{\tilde{x}\tilde{z}} + \frac{1}{2} u_{\tilde{x}}^2 u_{x\tilde{z}}\right), \nonumber \\
 H^3 = v(\tilde{x}u_{\tilde{x}} - \tilde{z}u_{\tilde{z}})u_{\tilde{x}\tilde{z}} - \frac{1}{2}\big(\tilde{x}u_{\tilde{x}}^2u_{x\tilde{z}} + \tilde{z}u_{\tilde{z}}^2u_{x\tilde{x}}\big), \nonumber \\
 H^4 = \left[v(u - xu_x) - \frac{t}{2}\big(v^2 + u_x^2\big)\right]u_{\tilde{x}\tilde{z}} + tu, \nonumber\\
 H^a = a(\tilde{x})\left\{u -\left[ vu_x + \frac{1}{2}\big(v^2 + u_x^2\big) \right]u_{\tilde{x}\tilde{z}}\right\}, \nonumber \\
 H^b = b(\tilde{z})\left\{\left[\frac{1}{2}\big(v^2 + u_x^2\big) - vu_x\right]u_{\tilde{x}\tilde{z}} - u\right\}, \nonumber\\
 H^{f,g} = vu_{\tilde{x}\tilde{z}}(f + g) + \frac{1}{2}\big[g_{\tilde{z}} u_xu_{\tilde{x}} - f_{\tilde{x}}u_xu_{\tilde{z}} + (f_t+g_t) u_{\tilde{x}} u_{\tilde{z}} \big],
 \label{int}
\end{gather}
with arbitrary smooth functions $a(\tilde{x})$, $b(\tilde{z})$, $f(t+x,\tilde{x})$, $g(t-x,\tilde{z})$. According to our previous remarks, the Hamiltonian $H_1$ of the f\/irst heavenly f\/low, obtained in \eqref{H1}, is also contained in the set~\eqref{int} as a linear combination $H_1 = (H^{b=1} - H^{a=1})/2$.

Since the symmetry generators $X_1$, $X_2$, $X_3$, $Y_a$, $Z_b$ commute with the generator $-\partial_t$ of the f\/irst heavenly f\/low, the corresponding Hamiltonian densities $H^1$, $H^2$, $H^3$, $H^a$, $Z^b$ are densities of integrals of motion for the f\/low~\eqref{2comp}, subject to suitable boundary conditions. Moreover, each~$H^i$ is the density of the integral of motion along all the symmetry f\/lows with characteristics~\eqref{chartab} that commute with the corresponding~$X_i$.

Converting the 2-component f\/irst heavenly system back to the f\/irst heavenly equation in the one-component form, one could obtain its integrals as reductions of the integrals for the system.

\section{Lax pairs and recursion operators}\label{recurs}

We start with the truncated Lax pair, with the spectral parameter $\lambda = 0$, for the f\/irst heavenly equation \eqref{heav1} (see \cite{fer})
\begin{gather*}
 L_1 = u_{\tilde{t}\tilde{z}}D_{\tilde{y}} - u_{\tilde{t}\tilde{y}}D_{\tilde{z}},\qquad
 L_2 = u_{\tilde{x}\tilde{z}}D_{\tilde{y}} - u_{\tilde{x}\tilde{y}}D_{\tilde{z}},%\label{Lax}
\end{gather*}
which commute on solutions of~\eqref{heav1}. In the new variables $t = \tilde{t} + \tilde{y}$, $x=\tilde{t} - \tilde{y}$, used in the two-component system~\eqref{2comp}, they become
\begin{gather}
 L_1 = (u_{t\tilde{z}}+u_{x\tilde{z}})(D_t-D_x) - (u_{tt}-u_{xx})D_{\tilde{z}},\nonumber\\
 L_2 = u_{\tilde{x}\tilde{z}}(D_t-D_x) - (u_{t\tilde{x}}-u_{x\tilde{x}})D_{\tilde{z}}.
\label{L_12}
\end{gather}

Let $\varphi$ be a symmetry characteristic \cite{olv} of the f\/irst heavenly equation \eqref{heav1} in a one component form, so that the Lie equation reads $u_\tau = \varphi$, where $\tau$ is the group parameter and independent variables do not transform under the group. The equation \eqref{heav1}, transformed to the new variables, has the form
\begin{gather}
(u_{tt}-u_{xx})u_{\tilde{x}\tilde{z}} - (u_{t\tilde{z}}+u_{x\tilde{z}})(u_{t\tilde{x}}-u_{x\tilde{x}}) = 1,
\label{modheav}
\end{gather}
with the truncated Lax pair for this equation given in \eqref{L_12}. Symmetry condition for equation~\eqref{modheav} reads
\begin{gather}
 u_{\tilde{x}\tilde{z}}(\varphi_{tt}-\varphi_{xx}) + (u_{tt}-u_{xx})\varphi_{\tilde{x}\tilde{z}}
 - (u_{t\tilde{x}}-u_{x\tilde{x}})(\varphi_{t\tilde{z}}+\varphi_{x\tilde{z}})\nonumber\\
 \qquad {}- (u_{t\tilde{z}}+u_{x\tilde{z}})(\varphi_{t\tilde{x}}-\varphi_{x\tilde{x}}) = 0. \label{sym}
\end{gather}
It can be expressed in terms of Lax operators \eqref{L_12} as
\begin{gather*}
(D_t+D_x)L_2\varphi - D_{\tilde{x}}L_1\varphi = 0
%\label{LaxSym}
\end{gather*}
and hereby have a two-dimensional divergence form. Hence it def\/ines a potential variable $\tilde{\varphi}$ by the equations
\begin{gather}
 \tilde{\varphi}_t + \tilde{\varphi}_x = L_1 \varphi,\qquad \tilde{\varphi}_{\tilde{x}} = L_2 \varphi.\label{pot}
\end{gather}
It is easy to check that the symmetry condition can also be put in the form
\begin{gather*}
L_2(D_t+D_x)\varphi - L_1D_{\tilde{x}}\varphi = 0.
%\label{SymLax}
\end{gather*}
This equation is satisf\/ied for the potential $\tilde{\varphi}$. Indeed, using its def\/inition \eqref{pot} we have
\begin{gather*}L_2(D_t+D_x)\tilde{\varphi} - L_1D_{\tilde{x}}\tilde{\varphi} = [L_2,L_1]\varphi = 0, \end{gather*}
since $L_1$ and $L_2$ commute on solutions of the equation~\eqref{modheav}. Thus, the potential of a symmetry is again a symmetry and we obtain a recursion relation for symmetries.

In an explicit form the def\/inition \eqref{pot} of the symmetry potential becomes
\begin{gather}
 \tilde{\varphi}_{\tilde{x}} = u_{\tilde{x}\tilde{z}}(\varphi_t-\varphi_x) - (v_{\tilde{x}}-u_{x\tilde{x}})\varphi_{\tilde{z}},\qquad
 \tilde{\varphi}_t + \tilde{\varphi}_x = (v_{\tilde{z}}+u_{x\tilde{z}})(\varphi_t-\varphi_x) - \tilde{Q}\varphi_{\tilde{z}},
\label{potexpl}
\end{gather}
where $v=u_t$, and using the def\/inition of $Q$ in \eqref{2comp}, we def\/ine
\begin{gather*}
 \tilde{Q} = Q - u_{xx} = \frac{1}{u_{\tilde{x}\tilde{z}}}\left\{(v_{\tilde{x}}-u_{x\tilde{x}})(v_{\tilde{z}}+u_{x\tilde{z}}) + 1\right\}.
%\label{Q}
\end{gather*}
For the two-component form \eqref{2comp} of our equation, we introduce the two-component symmetry characteristic $(\varphi,\psi)$, where $\psi=\varphi_t$ and the same for the potential of the symmetry $(\tilde{\varphi},\tilde{\psi})$ with $\tilde{\psi}=\tilde{\varphi}_t$, and use this notation in~\eqref{potexpl} with the result
\begin{gather}
 \tilde{\varphi}_{\tilde{x}} = u_{\tilde{x}\tilde{z}}(\psi-\varphi_x) - (v_{\tilde{x}}-u_{x\tilde{x}})\varphi_{\tilde{z}},\qquad
 \tilde{\psi} + \tilde{\varphi}_x = (v_{\tilde{z}}+u_{x\tilde{z}})(\psi-\varphi_x) - \tilde{Q}\varphi_{\tilde{z}}.
\label{pot2comp}
\end{gather}
We solve the f\/irst equation here for $\tilde{\varphi}$ in the form
\begin{gather*}\tilde{\varphi} = D_{\tilde{x}}^{-1}\{u_{\tilde{x}\tilde{z}}(\psi-\varphi_x) - (v_{\tilde{x}}-u_{x\tilde{x}})\varphi_{\tilde{z}}\},\end{gather*}
and use this in the second equation in \eqref{pot2comp} to obtain
\begin{gather*}\tilde{\psi} = - D_{\tilde{x}}^{-1}D_x\{u_{\tilde{x}\tilde{z}}(\psi-\varphi_x) - (v_{\tilde{x}}-u_{x\tilde{x}})\varphi_{\tilde{z}}\}
+ (v_{\tilde{z}}+u_{x\tilde{z}})(\psi-\varphi_x) - \tilde{Q}\varphi_{\tilde{z}}.\end{gather*}
Then we obtain our f\/irst recursion operator $R_1$ in $2\times 2$ matrix form which acts on two-component symmetry characteristics
\begin{gather*}
\left(\begin{matrix}
 \tilde{\varphi}\\ \tilde{\psi}
\end{matrix}\right) = R_1 \left(\begin{matrix}
 \varphi\\ \psi
\end{matrix}\right)
%\label{R1act}
\end{gather*}
with the following explicit form for $R_1$
\begin{gather*}
R_1 = \left(\begin{matrix}
- D_{\tilde{x}}^{-1}\{u_{\tilde{x}\tilde{z}}D_x + (v_{\tilde{x}}-u_{x\tilde{x}})D_{\tilde{z}}\} & D_{\tilde{x}}^{-1}u_{\tilde{x}\tilde{z}}\\
 R_1^{21} & - D_{\tilde{x}}^{-1}D_x u_{\tilde{x}\tilde{z}} + v_{\tilde{z}}+u_{x\tilde{z}}
\end{matrix}\right),
 %\label{R1}
\end{gather*}
where
\begin{gather*}
R_1^{21} = D_{\tilde{x}}^{-1}D_x\{u_{\tilde{x}\tilde{z}}D_x + (v_{\tilde{x}}-u_{x\tilde{x}})D_{\tilde{z}}\}
- (v_{\tilde{z}}+u_{x\tilde{z}})D_x - \tilde{Q}D_{\tilde{z}}.
% \label{R_121}
\end{gather*}

We also discover a second recursion operator by noticing the discrete symmetry possessed both by the f\/irst heavenly equation in the two-component form and its symmetry condition. For equation~\eqref{heav1} we choose among its discrete symmetries the symmetry of exchanging $\tilde{x}\leftrightarrow -\tilde{z}$ and $\tilde{t}\leftrightarrow\tilde{y}$, which is inherited by the two-component system~\eqref{2comp} in the form
\begin{gather}
\tilde{x}\leftrightarrow -\tilde{z},\qquad t\leftrightarrow t,\qquad x\leftrightarrow -x.
\label{discr}
\end{gather}
 We note that this discrete symmetry also leaves invariant the symmetry condition for the f\/irst heavenly equation \eqref{sym}. The alternative Lax pair for this equation becomes
\begin{gather*}
 L^1 = u_{\tilde{t}\tilde{y}}D_{\tilde{x}} - u_{\tilde{x}\tilde{y}}D_{\tilde{t}},\qquad
 L^2 = u_{\tilde{x}\tilde{z}}D_{\tilde{t}} - u_{\tilde{t}\tilde{z}}D_{\tilde{x}},
%\label{Laxsym}
\end{gather*}
and the symmetry condition is now expressed as
\begin{gather*}
D_{\tilde{y}}L^2\varphi + D_{\tilde{z}}L^1\varphi = 0\quad \iff \quad L^2D_{\tilde{y}}\varphi + L^1D_{\tilde{z}}\varphi = 0.
%\label{LaxSymm}
\end{gather*}
Repeating the reasoning that has led us to the f\/irst recursion operator, we discover that the symmetry potential which is determined by the conditions
\begin{gather*}
 \hat{\varphi} = D_{\tilde{z}}^{-1}L^2 \varphi,\qquad \hat{\varphi} = - D_{\tilde{y}}^{-1}L^1 \varphi
%\label{potsym}
\end{gather*}
also satisf\/ies the symmetry condition and hereby is also a symmetry and hence we have obtained another recursion for symmetries. The second recursion operator in a matrix $2\times 2$ form, obtained from~$R_1$ by the same discrete symmetry transformation, reads
\begin{gather}
R_2 = - \left(\begin{matrix}
D_{\tilde{z}}^{-1}\{u_{\tilde{x}\tilde{z}}D_x - (v_{\tilde{z}}+u_{x\tilde{z}})D_{\tilde{x}}\} & D_{\tilde{z}}^{-1}u_{\tilde{x}\tilde{z}}\\
 R_2^{21} & D_{\tilde{z}}^{-1}D_x u_{\tilde{x}\tilde{z}} + v_{\tilde{x}}-u_{x\tilde{x}}
\end{matrix}\right),
 %\label{R2}
\end{gather}
where we have used $D_{\tilde{y}} = D_t - D_x$ for a two-component representation and
\begin{gather*}
R_2^{21} = D_{\tilde{z}}^{-1}D_x\{u_{\tilde{x}\tilde{z}}D_x - (v_{\tilde{z}}+u_{x\tilde{z}})D_{\tilde{x}}\}
+ (v_{\tilde{x}}-u_{x\tilde{x}})D_x - \tilde{Q}D_{\tilde{x}}.
% \label{R_221}
\end{gather*}
We have added extra overall minus in $R_2$ to have the action of the discrete symmetry to be $R_1\leftrightarrow R_2$.

For future use, we will need the recursion operators either even or odd relative to the discrete symmetry, i.e.,
$R_+ = R_1 + R_2$ and $R_- = R_1 - R_2$. Joining both cases we introduce $R_\varepsilon = R_1 + \varepsilon R_2$ with the explicit form
\begin{gather}
 R_\varepsilon = \left(\begin{matrix}
 R_\varepsilon^{11} & \big(D_{\tilde{x}}^{-1}-\varepsilon D_{\tilde{z}}^{-1}\big)u_{\tilde{x}\tilde{z}} \vspace{1mm}\\
 R_\varepsilon^{21}
 & - \big(D_{\tilde{x}}^{-1}+\varepsilon D_{\tilde{z}}^{-1}\big)D_x u_{\tilde{x}\tilde{z}} + v_{\tilde{z}}+u_{x\tilde{z}}
 -\varepsilon (v_{\tilde{x}}-u_{x\tilde{x}})
\end{matrix} \right),
\label{Reps}
\end{gather}
where $\varepsilon = \pm$,
\begin{gather*}
 R_\varepsilon^{11} = - \big(D_{\tilde{x}}^{-1}+\varepsilon D_{\tilde{z}}^{-1}\big) u_{\tilde{x}\tilde{z}}D_x
 - D_{\tilde{x}}^{-1}(v_{\tilde{x}}-u_{x\tilde{x}})D_{\tilde{z}} + \varepsilon D_{\tilde{z}}^{-1}(v_{\tilde{z}}+u_{x\tilde{z}})D_{\tilde{x}}, \\
 R_\varepsilon^{21} = \big(D_{\tilde{x}}^{-1}-\varepsilon D_{\tilde{z}}^{-1}\big)D_x u_{\tilde{x}\tilde{z}}D_x
+ D_x\big\{D_{\tilde{x}}^{-1}(v_{\tilde{x}}-u_{x\tilde{x}})D_{\tilde{z}}+ \varepsilon D_{\tilde{z}}^{-1}(v_{\tilde{z}}+u_{x\tilde{z}})D_{\tilde{x}}\big\}\\
 %\label{Reps11}
 \hphantom{R_\varepsilon^{21} =}{}
- \{v_{\tilde{z}}+u_{x\tilde{z}} + \varepsilon (v_{\tilde{x}}-u_{x\tilde{x}})\}D_x - \tilde{Q}(D_{\tilde{z}}-\varepsilon D_{\tilde{x}}).
 \end{gather*}

\section[Bi-Hamiltonian representations of the f\/irst heavenly equation in two-component form]{Bi-Hamiltonian representations of the f\/irst heavenly equation\\ in two-component form}\label{bi-Hamilt}

Composing the recursion operator $R_1$ with the f\/irst Hamiltonian operator $J_0$ in \eqref{J0} we obtain a candidate for a second Hamiltonian operator $J_1$ for the two-component system
\begin{gather*}
J_1 = R_1J_0 = \left(\begin{matrix}
 - D_{\tilde{x}}^{-1} & - \left\{D_{\tilde{x}}^{-1}D_x - \frac{\displaystyle(v_{\tilde{z}}+u_{x\tilde{z}})}{\displaystyle u_{\tilde{x}\tilde{z}}}\right\}\\
 \left\{D_{\tilde{x}}^{-1}D_x - \frac{\displaystyle(v_{\tilde{z}}+u_{x\tilde{z}})}{\displaystyle u_{\tilde{x}\tilde{z}}}\right\} & J_1^{22}
\end{matrix}\right),
% \label{J1}
\end{gather*}
where
\begin{gather*}
 J_1^{22} = D_{\tilde{x}}^{-1}D_x^2 - \frac{1}{u_{\tilde{x}\tilde{z}}}D_{\tilde{z}}\frac{1}{u_{\tilde{x}\tilde{z}}}
- (v_{\tilde{z}}+u_{x\tilde{z}})D_x \frac{1}{u_{\tilde{x}\tilde{z}}} - \frac{1}{u_{\tilde{x}\tilde{z}}}D_x (v_{\tilde{z}}+u_{x\tilde{z}})\\
\hphantom{J_1^{22} =}{}
 + \frac{(v_{\tilde{z}}+u_{x\tilde{z}})}{u_{\tilde{x}\tilde{z}}}D_{\tilde{x}}\frac{(v_{\tilde{z}}+u_{x\tilde{z}})}{u_{\tilde{x}\tilde{z}}}.
 %\label{J1_22}
\end{gather*}
The operator $J_1$ is obviously skew symmetric, so the remaining task is to check Jacobi identities which will be considered in the next section.

Similarly, composing the recursion operator $R_2$ with the Hamiltonian operator $J_0$ we obtain a candidate for a third Hamiltonian operator $J_2$ for the two-component system, which can be also obtained directly by applying the discrete symmetry transformation \eqref{discr} to $J_1$
\begin{gather*}
J_2 = R_2J_0 = \left(\begin{matrix}
 D_{\tilde{z}}^{-1} & - \left\{D_{\tilde{z}}^{-1}D_x + \frac{\displaystyle(v_{\tilde{x}}-u_{x\tilde{x}})}{\displaystyle u_{\tilde{x}\tilde{z}}}\right\}\\
 \left\{D_{\tilde{z}}^{-1}D_x + \frac{\displaystyle(v_{\tilde{x}}-u_{x\tilde{x}})}{\displaystyle u_{\tilde{x}\tilde{z}}}\right\} & J_2^{22}
\end{matrix}\right),
% \label{J2}
\end{gather*}
where
\begin{gather*}
 J_2^{22} =-\left\{ D_{\tilde{z}}^{-1}D_x^2 - \frac{1}{u_{\tilde{x}\tilde{z}}}D_{\tilde{x}}\frac{1}{u_{\tilde{x}\tilde{z}}}
+ (v_{\tilde{x}}-u_{x\tilde{x}})D_x \frac{1}{u_{\tilde{x}\tilde{z}}} + \frac{1}{u_{\tilde{x}\tilde{z}}}D_x (v_{\tilde{x}}-u_{x\tilde{x}})\right.\nonumber\\
\left.\hphantom{J_2^{22} =}{} + \frac{(v_{\tilde{x}}-u_{x\tilde{x}})}{u_{\tilde{x}\tilde{z}}}D_{\tilde{z}}\frac{(v_{\tilde{x}}-u_{x\tilde{x}})}{u_{\tilde{x}\tilde{z}}}\right\}.
 %\label{J2_22}
\end{gather*}
The operator $J_2$ is also obviously skew symmetric, so it remains only to check the Jacobi identities.

However, using simple natural ansatzes, we failed to f\/ind Hamiltonian functions corresponding to the operators $J_1$ and $J_2$ such that acting by these
operators on the variational derivatives of the Hamiltonian functions we could generate the two-component f\/low~\eqref{2comp}. Therefore, we considered
operators $J_+$ and $J_-$ generated from $J_0$ by the recursion operators $R_+$ and $R_-$, respectively, which have the additional property of being either
even or odd under the discrete symmetry transformation \eqref{discr}. Using simultaneously both even and odd operators in the form
$J_\varepsilon = R_\varepsilon J_0 = J_1 + \varepsilon J_2$, where $\varepsilon = \pm$ and using the formula \eqref{Reps} for $R_\varepsilon$, we obtain the following result
\begin{gather*}
 J_\varepsilon = \left(\begin{matrix}
 - \big(D_{\tilde{x}}^{-1} - \varepsilon D_{\tilde{z}}^{-1}\big) & \begin{matrix} - \big(D_{\tilde{x}}^{-1} + \varepsilon D_{\tilde{z}}^{-1}\big)D_x\\
 {}\displaystyle+ \frac{\{v_{\tilde{z}}+u_{x\tilde{z}}-\varepsilon (v_{\tilde{x}}-u_{x\tilde{x}})\}}{u_{\tilde{x}\tilde{z}}}
 \end{matrix}
 \\ \begin{matrix} \big(D_{\tilde{x}}^{-1} + \varepsilon D_{\tilde{z}}^{-1}\big)D_x\\
{} -\displaystyle \frac{\{v_{\tilde{z}}+u_{x\tilde{z}}-\varepsilon (v_{\tilde{x}}-u_{x\tilde{x}})\}}{u_{\tilde{x}\tilde{z}}}
 \end{matrix} & J_\varepsilon^{22}
 \end{matrix}\right),
%\label{Jeps}
\end{gather*}
where
\begin{gather*}
 J_\varepsilon^{22} = \big(D_{\tilde{x}}^{-1} - \varepsilon D_{\tilde{z}}^{-1}\big)D_x^2 -
\{v_{\tilde{z}}+u_{x\tilde{z}}+\varepsilon (v_{\tilde{x}}-u_{x\tilde{x}})\}D_x\frac{1}{u_{\tilde{x}\tilde{z}}} \\
\hphantom{J_\varepsilon^{22} =}{} - \frac{1}{u_{\tilde{x}\tilde{z}}}D_x\{v_{\tilde{z}}+u_{x\tilde{z}}+\varepsilon (v_{\tilde{x}}-u_{x\tilde{x}})\}
- \frac{1}{u_{\tilde{x}\tilde{z}}}(D_{\tilde{z}}-\varepsilon D_{\tilde{x}})\frac{1}{u_{\tilde{x}\tilde{z}}}  \\
\hphantom{J_\varepsilon^{22} =}{} + \frac{(v_{\tilde{z}}+u_{x\tilde{z}})}{u_{\tilde{x}\tilde{z}}}D_{\tilde{x}} \frac{(v_{\tilde{z}}+u_{x\tilde{z}})}{u_{\tilde{x}\tilde{z}}}
-\varepsilon \frac{(v_{\tilde{x}}-u_{x\tilde{x}})}{u_{\tilde{x}\tilde{z}}}D_{\tilde{z}}\frac{(v_{\tilde{x}}-u_{x\tilde{x}})}{u_{\tilde{x}\tilde{z}}}.
 %\label{Jeps_22}
\end{gather*}
The operator $J_\varepsilon$ is also obviously skew symmetric. Jacobi identities for $J_\varepsilon$ and compatibility of it with $J_0$ and
also between $J_+$ and $J_-$ are proved in the next section, so that all of them are mutually compatible Hamiltonian operators.

For the operator $J_\varepsilon$ we have found the corresponding Hamiltonian densities $H_{0\varepsilon}$ using a~simple and natural ansatz of the density
to be linear in~$v$ and moreover
\begin{gather*}H_{0\varepsilon} = vh_{1\varepsilon}[u_{\tilde{x}\tilde{z}},u_{\tilde{x}},u_{\tilde{z}},\tilde{x},\tilde{z},x] + h_{0\varepsilon}[u],\end{gather*}
where $h_{0\varepsilon}[u]$ depends only on $u$ and its partial derivatives. Then the coef\/f\/icient functions $h_{1\varepsilon}$ and $h_{0\varepsilon}$
are determined from the requirement that the two-component system~\eqref{2comp} should admit two bi-Hamiltonian representations for $\varepsilon=\pm$
with the Hamiltonian operators $J_\varepsilon$
\begin{gather}
 \left(\begin{matrix}
 u_t\\ v_t
 \end{matrix}\right) = J_\varepsilon \left(\begin{matrix}
 \delta_u H_{0\varepsilon}\\ \delta_v H_{0\varepsilon}
 \end{matrix}\right) = \left(\begin{matrix}
 v\\ Q
 \end{matrix}\right),
\label{biHam}
\end{gather}
where operators $\delta_u$, $\delta_v$ denote Euler--Lagrange operators of the Hamiltonian density \cite{olv} equivalent to variational derivatives of the Hamiltonian functional $\mathscr{H} = \iiint\limits_{-\infty}^{+\infty}H_0 d\tilde{x} d\tilde{z} dx$ and $Q$ is def\/ined in~\eqref{2comp}. This requirement completely determines (up to a total divergence) the Hamiltonian densities to be
\begin{gather}
 H_{0\varepsilon} = \frac{1}{2} (\varepsilon\tilde{x} - \tilde{z})vu_{\tilde{x}\tilde{z}}- \frac{1}{4} u_x(u_{\tilde{x}} + \varepsilon u_{\tilde{z}})
\label{dens}
\end{gather}
and the equation \eqref{biHam} indeed yields the second and third Hamiltonian representations of the f\/irst heavenly equation in a two-component evolutionary form with the Hamiltonian densities~\eqref{dens} for $\varepsilon = \pm$. Together with the original Hamiltonian representation~\eqref{J0H1} of the system~\eqref{2comp} we end up with the tri-Hamiltonian representation of this system
\begin{gather}
 \left(\begin{matrix}
 u_t\\ v_t
 \end{matrix}\right) = J_0
 \left( \begin{matrix}
 \delta_uH_1\\ \delta_vH_1
 \end{matrix}\right) = J_+ \left(\begin{matrix}
 \delta_u H_{0+}\\ \delta_v H_{0+}
 \end{matrix}\right) =
 J_- \left(\begin{matrix}
 \delta_u H_{0-}\\ \delta_v H_{0-}
 \end{matrix}\right)
 \label{twobiHam}
 \end{gather}
provided that we have proved the Hamiltonian property for the operators $J_+$ and $J_-$ and compatibility of the Hamiltonian operators $J_0$, $J_+$ and $J_-$.
These operators were automatically Hamiltonian if we would know that the recursion operators $R_+$ and $R_-$ are hereditary (Nijenhuis) \cite{ff,sheftel}, but this property of our recursion operators is not known.

For this reason, in section \ref{Jacobi} we check that the three operators $J_0$, $J_+$ and $J_-$ are mutually compatible Hamiltonian operators (form a \textit{Poisson pencil}). For this purpose we use the technique of the functional multi-vectors
from P.~Olver's book~\cite{olv}.

\section[Proof of Jacobi identities and compatibility of Hamiltonian structures]{Proof of Jacobi identities and compatibility\\ of Hamiltonian structures}\label{Jacobi}

Compatibility of three Hamiltonian structures means that the linear combination of the three Hamiltonian operators with arbitrary constant coef\/f\/icients is also a Hamiltonian operator. Since~$J_0$,~$J_+$ and~$J_-$ are obviously skew symmetric, the remaining problem is to prove the Jacobi identities for the linear combination $J = aJ_+ + bJ_- + cJ_0$ of the three Hamiltonian ope\-ra\-tors with arbitrary constant coef\/f\/icients $a$, $b$ and $c$. Thus, we check simultaneously that~$J_+$ and~$J_-$ are indeed Hamiltonian operators, at $b=c=0$ and $a=c=0$ respectively, that the three Hamiltonian operators $J_+$, $J_-$ and $J_0$ are compatible and the system~\eqref{2comp} in the form~\eqref{twobiHam} is \emph{tri-Hamiltonian}. Since $J_+ = J_1 + J_2$ and $J_- = J_1 - J_2$, we obviously can equivalently change the def\/inition of $J$ to $J = aJ_1 + bJ_2 + cJ_0$ in our analysis of the Jacobi identities for $J$ to simplify the calculations.

To prove the Jacobi identities for $J$, we use the technique of P.~Olver's book \cite{olv}, so below we give a short summary of the notation and results from this book. Let $A^l$ be the vector space of $l$-component dif\/ferential functions that depend on independent and dependent variables of the problem and also on partial derivatives of the dependent variables up to some f\/ixed order. A linear operator $J\colon A^l\rightarrow A^l$ is called \textit{Hamiltonian} if its Poisson bracket $\{\mathscr{P},\mathscr{Q}\}=\int\delta \mathscr{P}\cdot J\delta \mathscr{Q} dx dy dz$ is \textit{skew-symmetric}
\begin{gather*}%\label{skew}
 \{\mathscr{P},\mathscr{Q}\} = - \{\mathscr{Q},\mathscr{P}\},
\end{gather*}
and satisf\/ies the \textit{Jacobi identity}
\begin{gather}\label{jacobi}
 \{\{\mathscr{P},\mathscr{Q}\},\mathscr{R}\} + \{\{\mathscr{R},\mathscr{P}\},\mathscr{Q}\} + \{\{\mathscr{Q},\mathscr{R}\},\mathscr{P}\} = 0
\end{gather}
for all functionals $\mathscr{P}$, $\mathscr{Q}$ and $\mathscr{R}$, where $\delta$ is the variational derivative. However, the direct verif\/ication of the Jacobi identity \eqref{jacobi} is a hopelessly complicated computational task. For this reason we will use P.~Olver's theory of the functional multi-vectors, in particular, his criterion:

\begin{Theorem}[\protect{\cite[Theorem 7.8]{olv}}]
Let $\mathscr{D}$ be a skew-adjoint $l\times l$ matrix differential operator and $\Theta=\frac{1}{2}\int(\omega^T\wedge\mathscr{D}\omega) dx dy dz$ the corresponding functional bi-vector. Then $\mathscr{D}$ is Hamiltonian if and only if
 \begin{gather}
\operatorname{pr} {\bf v}_{\mathscr{D}\omega}(\Theta) = 0,
 \label{criterion}
 \end{gather}
where $\operatorname{pr} {\bf v}_{\mathscr{D}\omega}$ is a prolonged evolutionary vector field with the characteristic $\mathscr{D}\omega$ defined by
\begin{gather}
\operatorname{pr} {\bf v}_{\mathscr{D}\omega} = \sum\limits_{i,J}D_J\bigg(\sum\limits_j\mathscr{D}_{ij}\omega^{j}\bigg)\frac{\partial}{\partial u_{J}^{i}}, \qquad
 J=0,x,\tilde{x},\tilde{z},x\tilde{x},x\tilde{z},\tilde{x}\tilde{z},\dots.
 \label{PrVdef}
\end{gather}
\end{Theorem}

Here $u^i_0 = u^i$ and in our case $i,j = 1,2$, while $u^1=u$ and $u^2=v$, and $\omega = (\omega^1,\omega^2) = (\eta,\theta)$ is a functional one-form corresponding to a ``uni-vector'' with the following property for the action of total derivatives $D_J(\omega^i)=\omega^i_J$. By the def\/inition of the space of functional multi-vectors, the integrals of total divergences in both $\Theta$ and $\operatorname{pr} {\bf v}_{\mathscr{D}\omega}(\Theta)$ always vanish, so that we can integrate by parts terms in the integrands, if needed. We note also that by def\/inition of the prolonged evolutionary vector f\/ield $\operatorname{pr} {\bf v}_{\mathscr{D}\omega}$, it commutes with total derivatives and annihilates uni-vectors
\begin{gather*}\operatorname{pr} {\bf v}_{\mathscr{D}\omega}\big(\omega^i\big) = 0.\end{gather*}

In the following it is convenient to introduce the short-hand notation $\mu = v_{\tilde{z}} + u_{x\tilde{z}}$, $\nu = v_{\tilde{x}} - u_{x\tilde{x}}$ and
$A = u_{\tilde{x}\tilde{z}}$, so that $2A_x = \mu_{\tilde{x}} - \nu_{\tilde{z}}$ and $2v_{\tilde{x}\tilde{z}} = \mu_{\tilde{x}} + \nu_{\tilde{z}}$.
To check the Jacobi identities for the operator $J = aJ_1 + bJ_2 + cJ_0$ we set $\mathscr{D} = J$, where
\begin{gather}
 J = \left(
 \begin{matrix}
 bD_{\tilde{z}}^{-1} - aD_{\tilde{x}}^{-1} &
 \begin{matrix}-\left\{\big(aD_{\tilde{x}}^{-1} + bD_{\tilde{z}}^{-1}\big)D_x \right. \\ \left. {}+ \frac{1}{A} (b\nu - a\mu - c) \right\}
 \end{matrix} \vspace{1mm}\\
 \begin{matrix}\left\{\big(aD_{\tilde{x}}^{-1} + bD_{\tilde{z}}^{-1}\big)D_x \right. \\ \left. {}+ \frac{1}{A} (b\nu - a\mu - c) \right\}
 \end{matrix} & J^{22}
 \end{matrix}
 \right),
 \label{J}
\end{gather}
where
\begin{gather}
 J^{22} = \big(aD_{\tilde{x}}^{-1} - bD_{\tilde{z}}^{-1}\big)D_x^2 + \frac{1}{A}\big(bD_{\tilde{x}}^{-1} - aD_{\tilde{z}}^{-1}\big)\frac{1}{A}
 - (a\mu + b\nu)D_x\frac{1}{A} \nonumber \\
 \hphantom{J^{22} =}{} - \frac{1}{A}D_x(a\mu + b\nu) + \frac{\mu}{A}D_{\tilde{x}}\frac{1}{A}(a\mu + c) - \frac{\nu}{A}D_{\tilde{z}}\frac{1}{A}(b\nu - c)
 + \frac{cv_{\tilde{x}\tilde{z}}}{A^2}.
 \label{J22}
\end{gather}
The bi-vector $\Theta$ in the theorem above has the form
\begin{gather}
 2\Theta = \int (\eta,\theta)\wedge \left(
 \begin{matrix}
 J^{11} & J^{12} \\
 J^{21} & J^{22}
 \end{matrix} \right)
 \left( \begin{matrix}
 \eta \\
 \theta
 \end{matrix}
 \right) d\tilde{x} d\tilde{z} dx \nonumber\\
\hphantom{2\Theta}{} = \int \Bigl\{ \eta\wedge \big(bD_{\tilde{z}}^{-1} - aD_{\tilde{x}}^{-1}\big)\eta - \eta\wedge \big(aD_{\tilde{x}}^{-1} + bD_{\tilde{z}}^{-1}\big)\theta_x \nonumber\\
\hphantom{2\Theta=}{} - 2\eta\wedge \frac{1}{A}(b\nu - a\mu - c)\theta + \theta\wedge \big(aD_{\tilde{x}}^{-1} + bD_{\tilde{z}}^{-1}\big)\eta_x
 \nonumber \\
 \hphantom{2\Theta=}{} + \theta\wedge \big(aD_{\tilde{x}}^{-1} - bD_{\tilde{z}}^{-1}\big)\theta_{xx} + \theta\wedge\frac{1}{A^2}(b\theta_{\tilde{x}}-a\theta_{\tilde{z}}) \nonumber \\
\hphantom{2\Theta=}{} - 2\theta\wedge\frac{1}{A}(a\mu + b\nu)\theta_x + \theta\wedge \frac{\mu}{A^2}(a\mu + c)\theta_{\tilde{x}}
 - \theta\wedge \frac{\nu}{A^2}(b\nu - c)\theta_{\tilde{z}} \Bigr\} d\tilde{x} d\tilde{z} dx .\label{Theta}
\end{gather}

According to P.~Olver's criterion \eqref{criterion} with $\mathscr{D} = J$ applied to $\Theta$ in \eqref{Theta}, the condition for Jacobi identities to be satisf\/ied reads
\begin{gather}
\operatorname{pr} {\bf v}_{J\omega}(\Theta) = 0,
\label{criterJ}
\end{gather}
where $\operatorname{pr} {\bf v}_{J\omega}$ is the prolonged evolutionary vector f\/ield with the characteristic $J\omega$ which acts on each term in the integrand of~\eqref{Theta}. We will further skip the integral sign in the condition~\eqref{criterJ}, keeping in mind the possibility of integration by parts while always omitting total divergences, and leaving out the characteristic $J\omega$ in the notation~$\operatorname{pr} {\bf v}$ for the prolonged vector f\/ield in~\eqref{criterJ} for brevity. The criterion~\eqref{criterJ} becomes
\begin{gather}
 - 2\eta\wedge\operatorname{pr} {\bf v} \left[\frac{1}{A}(b\nu-a\mu-c)\right]\wedge\theta + b\theta\wedge\operatorname{pr} {\bf v}\left(\frac{1}{A^2}\right)\wedge\theta_{\tilde{x}}
 \nonumber\\
\qquad{} - a\theta\wedge\operatorname{pr} {\bf v} \left(\frac{1}{A^2}\right)\wedge\theta_{\tilde{z}} - 2\theta\wedge\operatorname{pr} {\bf v} \left[\frac{(a\mu+b\nu)}{A}\right] \wedge\theta_x
\nonumber\\
\qquad{} + \theta\wedge\operatorname{pr} {\bf v} \left[\frac{\mu(a\mu+c)}{A^2}\right]\wedge\theta_{\tilde{x}}
- \theta\wedge\operatorname{pr} {\bf v} \left[\frac{\nu(b\nu-c)}{A^2}\right]\wedge\theta_{\tilde{z}} = 0\ \ \textrm{(mod DIV)},
\label{critexp}
\end{gather}
where ``mod DIV'' (skipped in the following) means that the left-hand side of \eqref{critexp} equal to a~total divergence is equivalent to zero.

{\allowdisplaybreaks
The action of the vector f\/ield $\operatorname{pr} {\bf v}_{J\omega}$ involved in~\eqref{critexp} is def\/ined in terms of the matrix elements of the operator $J$ in~\eqref{J}, \eqref{J22} according to the following rules (with $u=u^1$, $v=u^2$ and the letter subscripts denoting partial derivatives)
\begin{gather*}
 \operatorname{pr} {\bf v}_{J\omega}(A) = \operatorname{pr} {\bf v}_{J\omega}(u_{\tilde{x}\tilde{z}}) = D_{\tilde{x}}D_{\tilde{z}}\big\{J^{11}\eta + J^{12}\theta\big\} \nonumber\\
 \hphantom{\operatorname{pr} {\bf v}_{J\omega}(A)}{} = D_{\tilde{x}}D_{\tilde{z}}\left\{\big(bD_{\tilde{z}}^{-1} - aD_{\tilde{x}}^{-1}\big)\eta - \big(aD_{\tilde{x}}^{-1} + bD_{\tilde{z}}^{-1}\big)\theta_x
 - \frac{1}{A}(b\nu - a\mu - c)\theta\right\}\nonumber\\
 \hphantom{\operatorname{pr} {\bf v}_{J\omega}(A)}{}
 = b\eta_{\tilde{x}} - a\eta_{\tilde{z}} - a\theta_{x\tilde{z}} - b\theta_{x\tilde{x}}
 + \left[\frac{(a\mu - b\nu +c)}{A}\theta\right]_{\tilde{x}\tilde{z}},%\label{prA}
\\
 \operatorname{pr} {\bf v}_{J\omega}(\mu) = \operatorname{pr} {\bf v}_{J\omega}(v_{\tilde{z}} + u_{x\tilde{z}}) = D_{\tilde{z}} \big\{J^{21}\eta + J^{22}\theta + D_x\big(J^{11}\eta + J^{12}\theta\big)\big\} \\
\hphantom{\operatorname{pr} {\bf v}_{J\omega}(\mu)}{} = 2b\eta_x + \left[\frac{(b\nu-a\mu-c)}{A}\eta\right]_{\tilde{z}} - 2b\theta_{xx}
 + \left[\frac{b\theta_{\tilde{x}} - a\theta_{\tilde{z}}}{A^2}\right]_{\tilde{z}} - \left[\frac{(bA_{\tilde{x}} - aA_{\tilde{z}})}{A^3}\theta\right]_{\tilde{z}} \\
 \hphantom{\operatorname{pr} {\bf v}_{J\omega}(\mu)=}{}
 - \left[\frac{(a\mu+3b\nu-c)}{A}\theta_x\right]_{\tilde{z}}- \left\{\left[\frac{a\mu+b\nu}{A}\right]_x \theta\right\}_{\tilde{z}}
 + \left\{\frac{\mu}{A}\left[\frac{(a\mu+c)}{A}\theta\right]_{\tilde{x}}\right\}_{\tilde{z}} \nonumber \\
\hphantom{\operatorname{pr} {\bf v}_{J\omega}(\mu)=}{} - \left\{\frac{\nu}{A}\left[\frac{(b\nu-c)}{A}\theta\right]_{\tilde{z}}\right\}_{\tilde{z}}
 + \left\{\left[\frac{a\mu-b\nu+c}{A}\right]_x\theta\right\}_{\tilde{z}} + c\left(\frac{v_{\tilde{x}\tilde{z}}}{A^2}\theta\right)_{\tilde{z}},
%\label{prmu}
\\
 \operatorname{pr} {\bf v}_{J\omega}(\nu) = \operatorname{pr} {\bf v}_{J\omega}(v_{\tilde{x}} - u_{x\tilde{x}}) = D_{\tilde{x}} \big\{J^{21}\eta + J^{22}\theta - D_x\big(J^{11}\eta + J^{12}\theta\big)\big\} \nonumber \\
 \hphantom{\operatorname{pr} {\bf v}_{J\omega}(\nu)}{} = 2a\eta_x + \left[\frac{(b\nu-a\mu-c)}{A}\eta\right]_{\tilde{x}} + 2a\theta_{xx}
 + \left[\frac{b\theta_{\tilde{x}} - a\theta_{\tilde{z}}}{A^2}\right]_{\tilde{x}} - \left[\frac{(bA_{\tilde{x}} - aA_{\tilde{z}})}{A^3}\theta\right]_{\tilde{x}} \\
 \hphantom{\operatorname{pr} {\bf v}_{J\omega}(\nu)=}{}
 - \left[\frac{(3a\mu+b\nu+c)}{A}\theta_x\right]_{\tilde{x}}- \left\{\left[\frac{a\mu+b\nu}{A}\right]_x \theta\right\}_{\tilde{x}}
 + \left\{\frac{\mu}{A}\left[\frac{(a\mu+c)}{A}\theta\right]_{\tilde{x}}\right\}_{\tilde{x}} \\
\hphantom{\operatorname{pr} {\bf v}_{J\omega}(\nu)=}{}
- \left\{\frac{\nu}{A}\left[\frac{(b\nu-c)}{A}\theta\right]_{\tilde{z}}\right\}_{\tilde{x}}
 - \left\{\left[\frac{a\mu-b\nu+c}{A}\right]_x\theta\right\}_{\tilde{x}} + c\left(\frac{v_{\tilde{x}\tilde{z}}}{A^2}\theta\right)_{\tilde{x}}.
%\label{prnu}
\end{gather*}}

With these prolongation formulas plugged in the criterion \eqref{critexp}, three groups of terms, bilinear in $\eta$ and its derivatives, linear in $\eta$ and without $\eta$, should separately either vanish or be reducible to a total divergence form using integration by parts when necessary. The operator
$J = aJ_1 + bJ_2 + cJ_0$ should satisfy the Jacobi identities for arbitrary parameters $a$, $b$ and $c$ and hence vanishing of terms, up to a total divergence, should occur separately in each subgroup with distinct bilinear dependence on the coef\/f\/icients $a$, $b$ and $c$. Moreover, each subgroup is further divided into subgroups of terms with the trilinear, bilinear, linear dependence on $\mu$, $\nu$ and its derivatives or terms without $\mu$ and $\nu$ and each such subgroup also should separately vanish or be reducible to a total divergence. Sometimes such cancelations of terms will occur only on account of the identities $2A_x = \mu_{\tilde{x}} - \nu_{\tilde{z}}$ and $2v_{\tilde{x}\tilde{z}} = \mu_{\tilde{x}} + \nu_{\tilde{z}}$ following from def\/initions of $\mu$, $\nu$ and $A$ given before the formula \eqref{J}.

It is easy to check that all terms bilinear in $\eta$ and its derivatives are canceled. For terms linear in $\eta$ we consider separately groups containing
$a^2$, $b^2$, $c^2$, $ac$, $bc$ and $ab$. For example, terms with $a^2$ without $\mu$ (there are no terms with $\nu$ in this group) are combined into the single term (we set $a=1$ here)
\begin{gather*}-2\left(\eta\wedge\frac{\theta_{\tilde{z}}}{A^3}\right)_{\tilde{z}}\wedge\theta = -2\left(\eta\wedge\frac{\theta_{\tilde{z}}}{A^3}\wedge\theta\right)_{\tilde{z}},\end{gather*}
which is a total divergence. Terms linear in $\mu$ combine to
\begin{gather*}-\frac{2}{A}\left(\eta\wedge\frac{\mu}{A}\theta_x\right)_{\tilde{z}}\wedge\theta + \frac{2}{A}\left(\frac{\mu}{A}\eta\wedge\theta_x\right)_{\tilde{z}}\wedge\theta = 0.\end{gather*}
Terms quadratic in $\mu$ are{\samepage
\begin{gather*}
\frac{2}{A}\eta\wedge\left\{\left(\frac{\mu^2}{A^2}\theta_{\tilde{x}}\right)_{\tilde{z}} + \frac{\mu}{A}\left(\frac{\mu}{A}\right)_{\tilde{x}}\theta_{\tilde{z}} \right\}\wedge\theta
- \frac{2\mu}{A^2}\eta\wedge\left\{\left(\frac{\mu}{A}\theta_{\tilde{z}}\right)_{\tilde{x}} + \left(\frac{\mu}{A}\right)_{\tilde{z}}\theta_{\tilde{x}}\right\}\wedge\theta \\
\qquad {}- \frac{2\mu}{A^2}\theta\wedge\left(\frac{\mu}{A}\eta\right)_{\tilde{z}}\wedge\theta_{\tilde{x}} + \frac{2\mu^2}{A^3}\theta\wedge\eta_{\tilde{z}}\wedge\theta_{\tilde{x}} = 0,
\end{gather*}
where the cancelations are obvious after expanding all derivatives of the products.}

As another example, we consider terms linear in $\eta$ containing $ab$. Terms without $\mu$ and $\nu$ do not cancel completely, the remainder being (we skip $ab$ in the following)
\begin{gather*}- \frac{8A_x}{A^2}\eta\wedge\theta_x\wedge\theta = - \frac{4}{A^2}(\mu_{\tilde{x}} - \nu_{\tilde{z}})\eta\wedge\theta_x\wedge\theta,\end{gather*}
which should be joined with the terms linear in $\mu$ and $\nu$. All such terms f\/inally combine to the following expressions
\begin{gather*}
4\left(\frac{\mu}{A^2}\eta\wedge\theta_{\tilde{x}}\right)_x \wedge\theta + 4\left(\frac{\mu}{A^2}\eta\wedge\theta_x\right)_{\tilde{x}} \wedge\theta
- 4\left(\frac{\nu}{A^2}\eta\wedge\theta_{\tilde{z}}\right)_x \wedge\theta \\
\qquad\quad{} - 4\left(\frac{\nu}{A^2}\eta\wedge\theta_x\right)_{\tilde{z}} \wedge\theta + \frac{4A_x}{A^3} \eta\wedge(\mu\theta_{\tilde{x}}- \nu\theta_{\tilde{z}})\wedge\theta \\
\qquad{} = - 4\frac{\mu}{A^2}\eta\wedge\theta_{\tilde{x}}\wedge\theta_x - 4\frac{\mu}{A^2}\eta\wedge\theta_x\wedge\theta_{\tilde{x}}
+ 4\frac{\nu}{A^2}\eta\wedge\theta_{\tilde{z}}\wedge\theta_x + 4\frac{\nu}{A^2}\eta\wedge\theta_x\wedge\theta_{\tilde{z}}\\
\qquad\quad{} + \frac{4A_x}{A^3} \eta\wedge(\mu\theta_{\tilde{x}}- \nu\theta_{\tilde{z}})\wedge\theta = \frac{2\mu(\mu_{\tilde{x}}-\nu_{\tilde{z}})}{A^3} \eta\wedge \theta_{\tilde{x}}\wedge\theta - \frac{2\nu(\mu_{\tilde{x}}-\nu_{\tilde{z}})}{A^3} \eta\wedge\theta_{\tilde{z}}\wedge\theta,
\end{gather*}
where we have used integration by parts skipping total divergencies and, at the large step, the relation $2A_x = \mu_{\tilde{x}} - \nu_{\tilde{z}}$.
The remaining term is bilinear in $\mu$ and $\nu$ and it should be joined to other bilinear terms, linear in $\eta$ and containing $ab$, which jointly yield
\begin{gather*}- \left(\frac{2\mu^2}{A^3}\eta\wedge\theta_{\tilde{x}}\right)_{\tilde{x}} \wedge\theta
- \left(\frac{2\nu^2}{A^3}\eta\wedge\theta_{\tilde{z}}\right)_{\tilde{z}} \wedge\theta
+ \left(\frac{2\mu\nu}{A^3}\eta\wedge\theta_{\tilde{x}}\right)_{\tilde{z}} \wedge\theta
+ \left(\frac{2\mu\nu}{A^3}\eta\wedge\theta_{\tilde{z}}\right)_{\tilde{x}} \wedge\theta = 0,\end{gather*}
which either vanish or cancel after integrating by parts.

In a similar way, we check the cancelation of linear in $\eta$ terms proportional to $b^2$, $c^2$, $ac$ and $bc$ and also distinct cancelation of groups of terms without $\eta$ that contain each of bilinear combinations of the coef\/f\/icients $a$, $b$ and $c$. Each such group is subdivided into subgroups with dif\/ferent powers of $\mu$, $\nu$ and their derivatives. The calculations are straightforward but much too lengthy to be presented here, especially for the groups of terms without $\eta$.

Thus, the Jacobi identities are satisf\/ied for the linear combination $J = aJ_1 + bJ_2 + cJ_0$ and equivalently for $J = aJ_+ + bJ_- + cJ_0$ with arbitrary constant coef\/f\/icients which proves that all the three operators $J_+$, $J_-$ and $J_0$ are Hamiltonian and mutually compatible as soon as P.~Olver's criterion is applicable. However, P.~Olver's criterion is formulated for matrix-dif\/ferential operators, while our operators $J_+$ and $J_-$ are nonlocal. An extensive literature exists on the theory of nonlocal Hamiltonian operators in $1+1$ dimensions, e.g.,~\cite{sk,fe,fe+,mo,se}. However, to the authors' knowledge, no ef\/fective methods exist for checking the Jacobi identities in the multi-dimensional case except O.~Mokhov's method in~\cite{mo98} whose ef\/f\/iciency depends on the level of complexity of the nonlocality. As a hint to the applicability of P.~Olver's method for nonlocal Hamiltonian operators, we can regard the papers~\cite{olv2, osw}, where the Jacobi identities for nonlocal symmetries were shown to be correct if nonlocal variables are included in symmetries' characteristics, which imply ``ghost'' terms in the commutators. Since according to P.~Olver's method nonlocal terms are automatically included in the characteristic of the evolutionary vector f\/ield $\operatorname{pr} {\bf v}_{J\omega}$, with nonlocal $J$, and all such terms are canceled in the process of application of the criterion, we believe that the criterion works correctly in this more general case, which we consider as a well-founded conjecture. Thus, under this conjecture we have shown that the f\/irst heavenly equation in the two component evolutionary form~\eqref{2comp}, being converted to the form~\eqref{twobiHam}, is indeed a tri-Hamiltonian system.

Of course, a rigorous formulation of this method for checking the Jacobi identities for nonlocal Hamiltonian operators is still awaited and could be a very worthwhile project \cite{Olver}.

\section{Conclusion}

\looseness=-1
We have converted the f\/irst heavenly equation into a two-component evolutionary form. Using two dif\/ferent Lax pairs derived for the one-component form of this equation, we have constructed two independent recursion operators $R_1$ and $R_2$ for symmetries of the f\/irst heavenly system (FHS) from the Lax operators. They are related by a discrete symmetry of both the FHS and its symmetry condition. We have obtained a Lagrangian for the two-component FHS. Applying to this degenerate Lagrangian the Dirac's theory of constraints, we have obtained a symplectic operator and its inverse, the latter being a Hamiltonian operator $J_0$. We have found the corresponding Hamiltonian density, thus presenting the f\/irst heavenly system in a~Hamiltonian form.

We have determined all local Lie point symmetries of the FHS system and, using the inverse Noether theorem in Hamiltonian form, we obtained Hamiltonians generating all the variational (Noether) point symmetries. These Hamiltonians are integrals of the motion along the f\/irst heavenly f\/low if the symmetry f\/lows generated by them commute with the FHS f\/low. Each Hamiltonian generating a variational point symmetry f\/low is conserved along each point symmetry f\/low that commutes with the f\/low generated by the Hamiltonian under study. Converting the FH system back to the f\/irst heavenly equation in the one-component form, one could obtain its integrals as reductions of the integrals for the system.

Composing each of the recursion operators with $J_0$, we have obtained candidates for two more Hamiltonian operators $J_1=R_1J_0$ and $J_2=R_2J_0$ which are related by the discrete symmetry transformation between the recursion operators. However, using some simple and natural ansatzes for densities we failed to f\/ind the Hamiltonian densities corresponding to $J_1$ and $J_2$, so that these operators cannot be used to convert the f\/irst heavenly f\/low in new Hamiltonian forms. Therefore, we introduced operators $J_+ = J_1+J_2$ and $J_- = J_1-J_2$, even and odd with respect to the discrete symmetry, respectively. For these operators we have succeeded to f\/ind corresponding Hamiltonian densities $H_{0+}$ and $H_{0-}$ by using a simple natural ansatz and hence we have obtained two more Hamiltonian representations for the f\/irst heavenly f\/low, provided that we prove that~$J_+$ and~$J_-$ are indeed Hamiltonian operators. These operators were automatically Hamiltonian if we would know that the recursion operators are hereditary (Nijenhuis), but this property of our recursion operators is not known. Therefore, since ope\-ra\-tors~$J_0$,~$J_+$ and $J_-$ are obviously skew-symmetric, the remaining problem is to check directly the Jacobi identities for these operators and prove the compatibility of the three Hamiltonian opera\-tors~$J_0$,~$J_+$ and~$J_-$. For this purpose we checked the Jacobi identities for the linear combination of these operators with arbitrary constant coef\/f\/icients. To prove the Jacobi identities directly appears to be ``a~hopelessly complicated computational task''~\cite{olv}. Therefore, we have applied P.~Olver's theory of functional multi-vectors from his book~\cite{olv}. In this way, under the well-founded conjecture of applicability of the Olver's method to nonlocal Hamiltonian operators, we have proved that all three operators are Hamiltonian (which has been already proved independently for $J_0$) and that all these operators are compatible. Hence, our main result is that the f\/irst heavenly equation in an evolutionary two-component form is a tri-Hamiltonian system. Earlier examples of tri-Hamiltonian systems, bidirectional Boussinesq system and three mutually compatible Hamiltonian structures associated with the Korteweg--de~Vries and Camassa--Holm bi-Hamiltonian f\/lows, are mentioned in P.~Olver's talk \cite{OlvTalk}. There are also many other well-known and famous multi-Hamiltonian systems (see, e.g., \cite{AF87,AF88,AF89,Kuper}).
The work on symmetries and conservation laws following from these Hamiltonian properties is now in progress.

\subsection*{Acknowledgements}

The research of M.B.~Sheftel is partly supported by the research grant from Bo\u{g}azi\c{c}i University Scientif\/ic Research Fund (BAP), research project No.~11643. The authors are thankful to the referees for their important remarks.

\pdfbookmark[1]{References}{ref}
\LastPageEnding

\end{document}